\documentclass[prb,onecolumn,12pt]{revtex4}
\usepackage{epsfig}
\usepackage{amssymb}
\usepackage{threeparttable}
\usepackage{graphicx}
\usepackage{longtable}

\begin{document}

\title{Low lattice thermal conductivity of stanene}
\author{Bo Peng$^1$, Hao Zhang$^{1,*}$, Hezhu Shao$^2$, Yuchen Xu$^1$, Xiangchao Zhang$^1$ and Heyuan Zhu$^1$}
\affiliation{$^1$Shanghai Ultra-precision Optical Manufacturing Engineering Center, Department of Optical Science and Engineering, Fudan University, Shanghai 200433, China\\
$^2$Ningbo Institute of Materials Technology and Engineering, Chinese Academy of Sciences, Ningbo 315201, China}

\begin{abstract}
A fundamental understanding of phonon transport in stanene is crucial to predict the thermal performance in potential stanene-based devices. By combining first-principle calculation and phonon Boltzmann transport equation, we obtain the lattice thermal conductivity of stanene. A much lower thermal conductivity (11.6 W/mK) is observed in stanene, which indicates higher thermoelectric efficiency over other 2D materials. The contributions of acoustic and optical phonons to the lattice thermal conductivity are evaluated. Detailed analysis of phase space for three-phonon processes shows that phonon scattering channels LA+LA/TA/ZA$\leftrightarrow$TA/ZA are restricted, leading to the dominant contributions of high-group-velocity LA phonons to the thermal conductivity. The size dependence of thermal conductivity is investigated as well for the purpose of the design of thermoelectric nanostructures.
\end{abstract}

\maketitle

\section{Introduction}

Research in 2D materials has experienced an explosion of interest due to their potential for integration into next-generation electronic and energy conversion devices \cite{C4NR01600A,Novoselov2012,Photothermoelectric-MoS2,Klinovaja2013}. Recently, a new 2D material, stanene, has been realized on the Bi$_2$Te$_3$(111) substrate by molecular beam epitaxy \cite{Zhu2015}, and attracted tremendous interest due to its extraordinary properties. Stanene, a monolayer of tin film with buckled honeycomb structure like silicene, is predicted to be an example of a topological insulator, which can support a large-gap 2D quantum spin Hall (QSH) state \cite{Matusalem2015,Liu2011,Liu2011a,stanene}. It has been reported that a nontrivial bulk gap of $\sim$0.1 eV opens when the spin-orbital coupling (SOC) is present in stanene, and the edge states are gapless with bands dispersing inside the bulk gap and helical with spin-momentum locking, which can be used as dissipationless conducting ``wires'' for electronic circuits \cite{Liu2011,stanene}. The free standing stanene with $\sim$0.1 eV SOC gap is a promising applicant for room temperature and even high temperature application. In addition, a recent study has found that the thermoelectric (TE) figure of merit $zT$ is strongly size dependent in stanene, and can be improved to be significantly larger than 1 by optimizing the geometric size, which provides great opportunities in the next-generation TE devices \cite{Xu2014a}.

All these electrical and thermoelectrical applications of stanene are closely related to its thermal transport properties. For instance, high TE performance requires the system to be a bad conductor for phonons but a good conductor for electrons, which often conflict with each other \cite{Qin2014,raey}. In 2D QSH systems such as stanene, the edge states form perfectly conducting channels for electrons, leading to great improvement in the TE figure of merit $zT$ \cite{Takahashi2010,1742-6596-334-1-012013,Xu2014a}. Moreover, $zT$ of stanene can be further improved by optimizing the geometry size to maximize the contribution of the gapless edge states, which significantly helps to optimize the electron transport. However, the phonon transport properties of stanene have not been investigated in detail. Furthermore, although the size dependence of $zT$ and the Seebeck coefficient in stanene has been studied for TE applications, the size dependence of lattice thermal conductivity, which is also crucial when designing TE nanostructures, remains uninvestigated.

Generally both electrons and phonons contribute to the total thermal conductivity. In this paper only the lattice thermal conductivity of stanene is under consideration. We calculate the lattice thermal conductivity $\kappa$ of stanene using first-principle calculation and an iterative solution of the Boltzmann transport equation (BTE) for phonons \cite{ShengBTE}. A much lower thermal conductivity (11.6 W/mK) is observed in stanene compared to other 2D materials, which makes stanene a promising candidate to realize high thermoelectric performance. We also extract the contribution of each phonon mode to investigate the underlying mechanism behind the novel thermal transport properties of stanene, and find that LA phonons contribute most to the thermal conductivity ($>$57\%) over a wide temperature range. It can be attributed to their high phonon group velocities and restricted phase space for three-phonon processes. Finally, we obtain the representative mean free path (MFP) of stanene that is useful when designing TE nanostructures. The MFP in the small-grain limit is also obtained when the nanostructuring induced phonon scattering dominates. The role of boundary scattering in stanene nanowires is examined as well.

\begin{figure}[ht]
\centering
\includegraphics[width=0.65\linewidth]{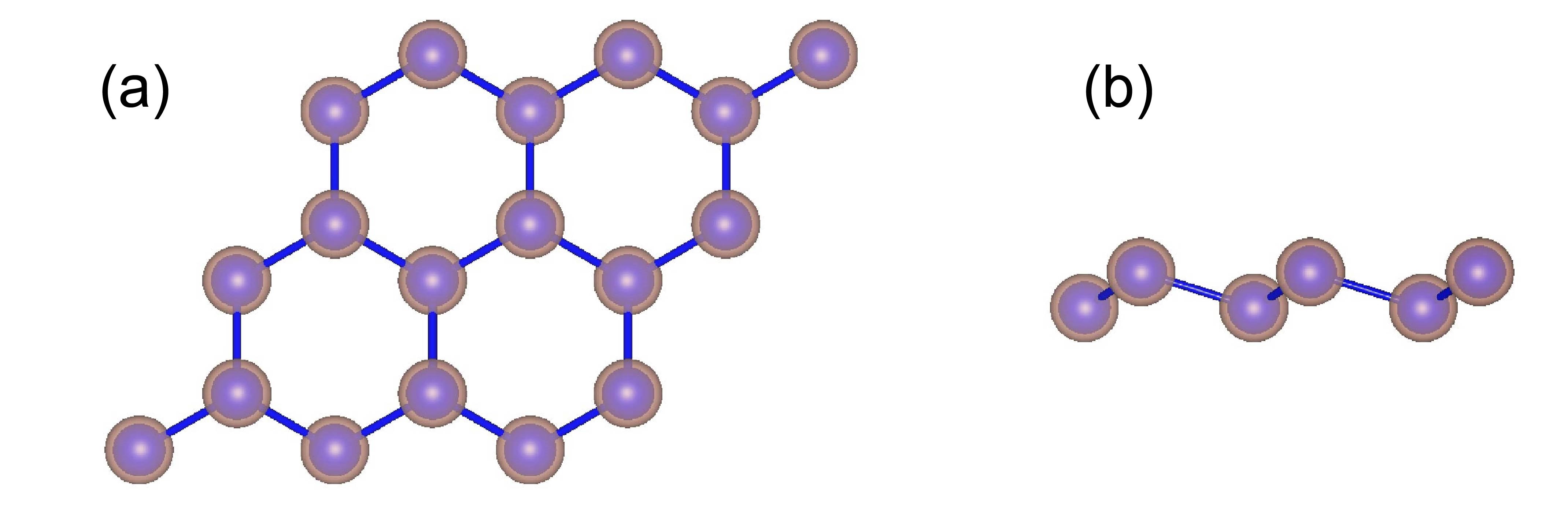}
\caption{(a)Top view and (b)side view of stanene.}
\label{lattice} 
\end{figure}

\begin{figure}[ht]
\centering
\includegraphics[width=0.5\linewidth]{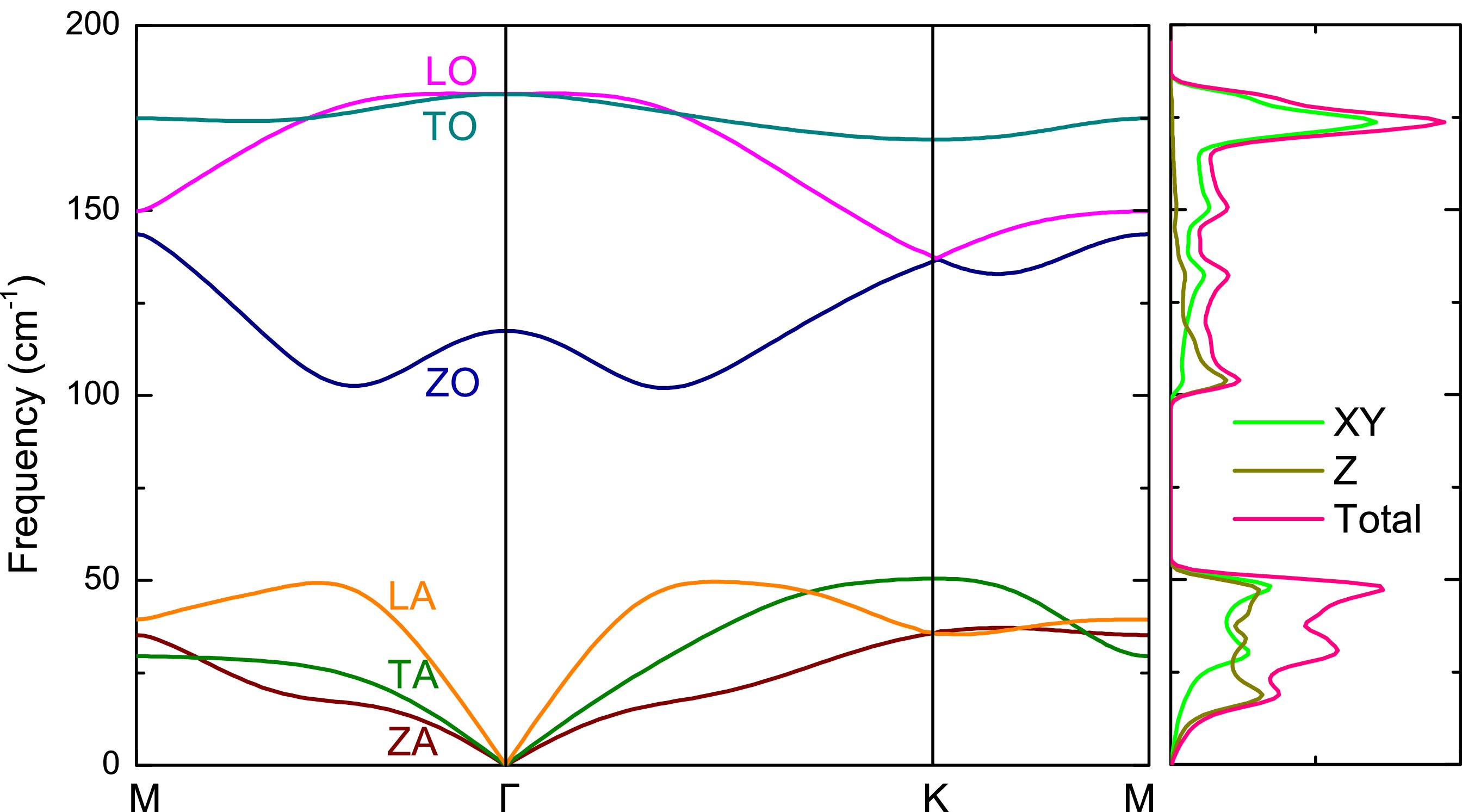}
\caption{Phonon dispersion and DOS of stanene. Three acoustic and three optical phonon modes are indicated with different colors.}
\label{phonon} 
\end{figure}

\section{Results and Discussions}
\subsection{Crystal structure and phonon dispersion}

Fig.~\ref{lattice} shows the optimized structure of stanene. A similar low-buckled configuration is also found in silicene, germanene and blue phosphorene \cite{Liu2011a,Matusalem2015,Liu2011,silicene,blue-phosphorene}, which is different from the planar geometry of graphene \cite{graphene}. The optimized lattice constant and buckling height are 4.67 \AA\ and 0.85 \AA, respectively, which are in good agreement with the previous work \cite{stanene,Matusalem2015,Liu2011}. There are two tin atoms in one unit cell, corresponding to three acoustic and three optical phonon branches.

Fig.~\ref{phonon} shows the calculated phonon dispersion. Similar to $\alpha$-Sn and $\beta$-Sn \cite{Tin}, the longitudinal acoustic (LA), transverse acoustic (TA), and z-direction acoustic (ZA) branches of stanene are linear near the $\Gamma$ point, which is distinctly different from the quadratic ZA-branch dispersion in graphene \cite{graphene-phonon}. This difference between graphene and stanene comes from structural difference. In planar graphene, the quadratic ZA-branch dispersion near the $\Gamma$ point can be explained as the consequence of the complete decoupling of the Z modes and XY modes due to the $D_{6h}$ point-group symmetry \cite{ZA,decoupling}. However, the buckled structure of stanene leads to the point group symmetry of $D_{3d}$, which breaks the reflection symmetry ($z\rightarrow -z$), and therefore the Z modes of stanene couple with the XY modes \cite{zengzhi}, which results in the linear dispersion of ZA phonons near the $\Gamma$ point.

\begin{figure}[ht]
\centering
\includegraphics[width=0.45\linewidth]{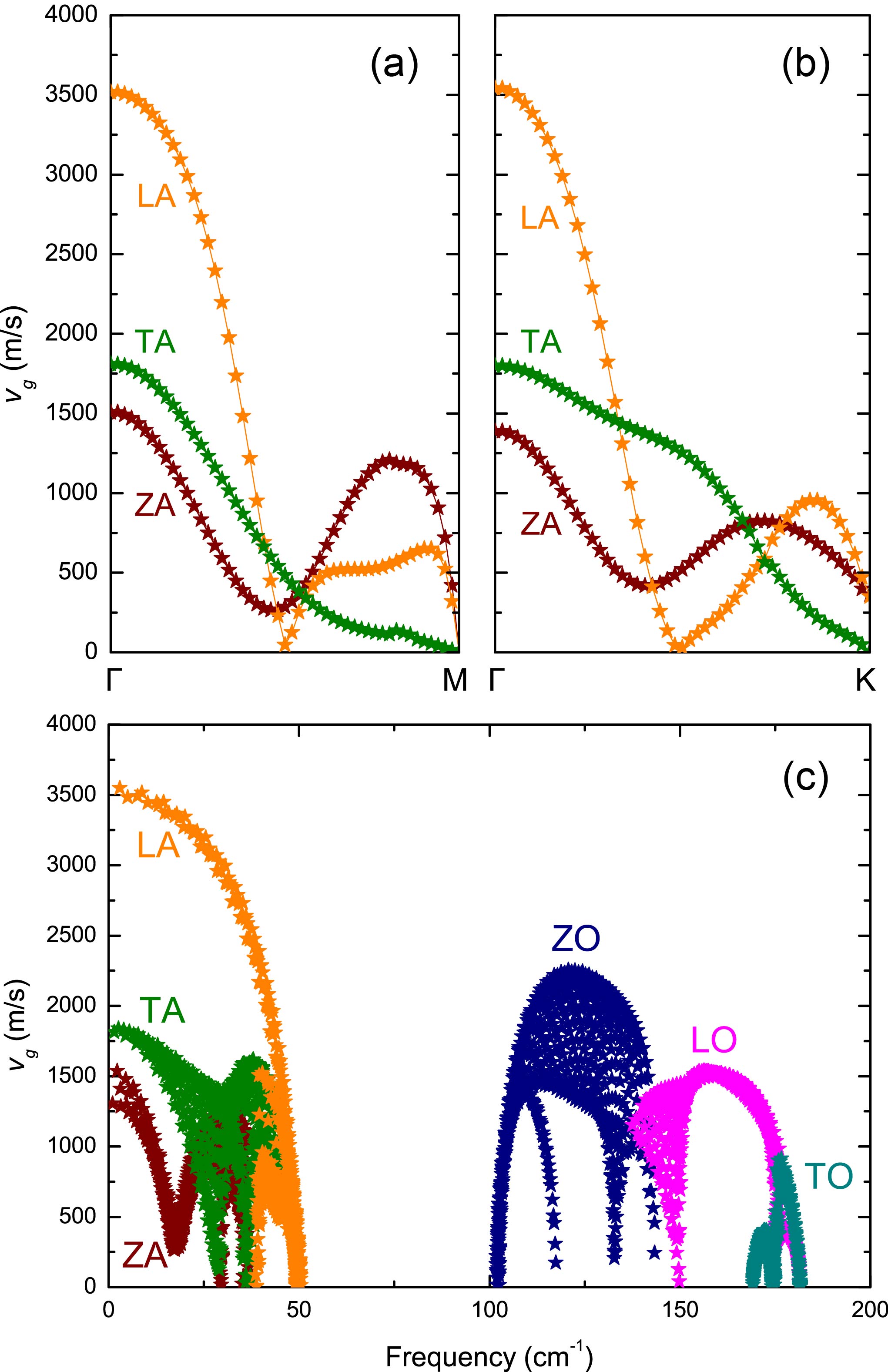}
\caption{Phonon group velocities along the (a)$\Gamma$-M and (b)$\Gamma$-K direction, and (c)within the first BZ.}
\label{velocities} 
\end{figure}

Using the phonon spectrum, we calculate the group velocities of acoustic phonons along $\Gamma$-M and $\Gamma$-K respectively, as shown in Fig.~\ref{velocities}(a) and (b). The anisotropy in group velocities along the two directions is weaker than that of black phosphorene and SnSe \cite{BP-phonon,SnSe}. As shown in Fig.~\ref{velocities}(c), the phonon group velocities indicate weak anisotropy within the entire first Brillouin zone (BZ). Much smaller sound velocities  are found in stanene (1300-3600 m/s compared to 5,400-8,800 m/s in silicene \cite{Li2013}, 4200-6,800 m/s in MoS$_2$ \cite{Liu2014-apl}, 4,000-8,000 m/s in blue phosphorene \cite{blue-phosphorene}, and 3,700-6,000 m/s in graphene \cite{Ong2011}). For LA, TA and ZA phonons, the sound velocities at long-wavelength limit are about 3548.7 m/s, 1811.4 m/s and 1303.7-1536.8 m/s, respectively.

\begin{figure}[ht]
\centering
\includegraphics[width=0.45\linewidth]{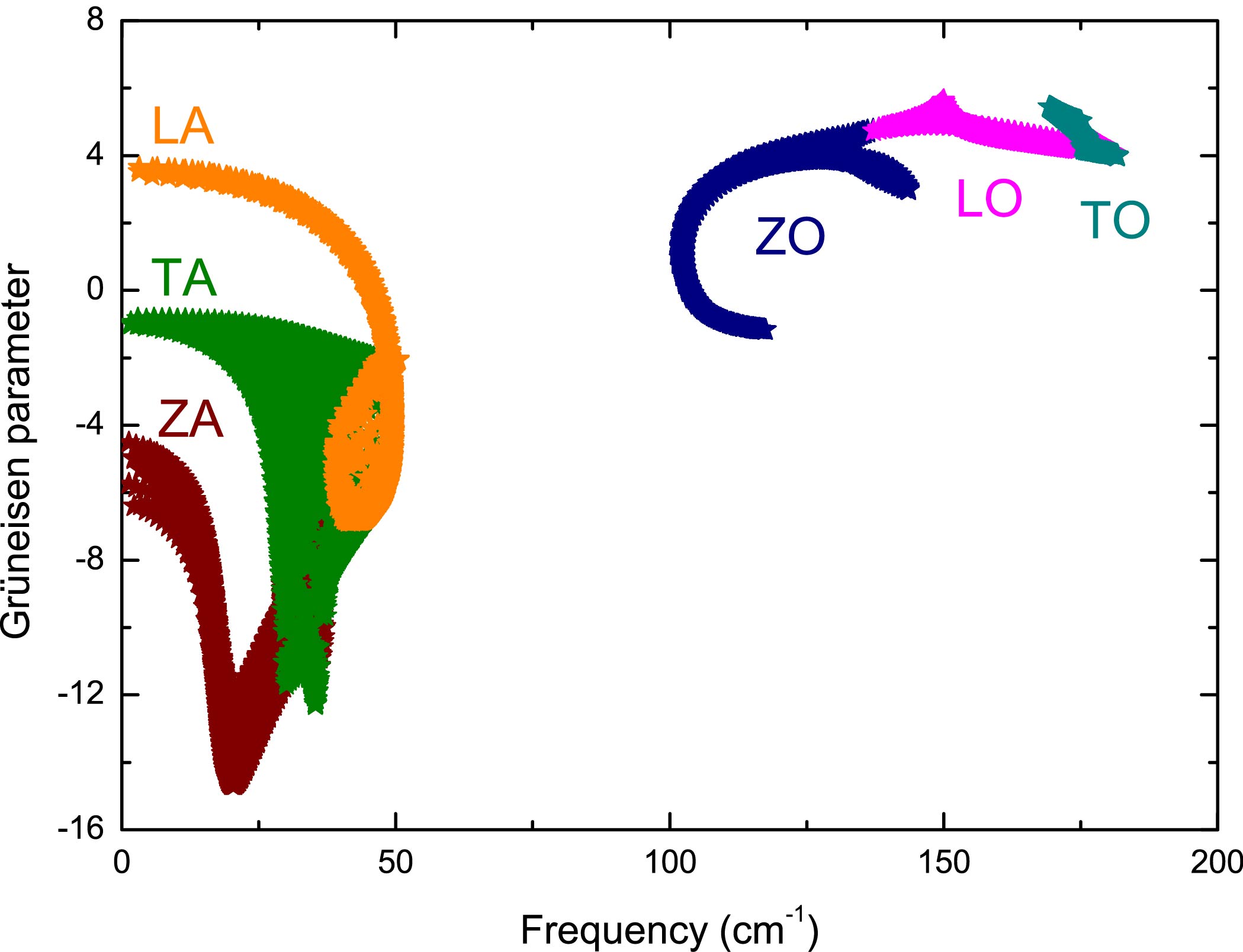}
\caption{Calculated Gr\"uneisen parameters within the first BZ.}
\label{gruneisen} 
\end{figure}

The Gr\"uneisen parameter , which provides information on the anharmonic interactions, is given by

\begin{equation}
\gamma_j(\textbf{q})=-\frac{a_0}{\omega_j(\textbf{q})}\frac{\partial\omega_j(\textbf{q})}{\partial a}
\end{equation}

\noindent where $a_0$ is the equilibrium lattice constant, $j$ is phonon branch index, and \textbf{q} is wave vector. Fig.~\ref{gruneisen} shows the Gr\"uneisen parameters within the first BZ. Strong anisotropy is found in the Gr\"uneisen parameters of TA phonons at high frequencies, while the isotropy in the Gr\"uneisen parameters of LA and TA phonons at low frequencies indicates that the anharmonicity of these phonons is similar from one phonon wave vector to another. The low-frequency phonons contribute most to the lattice thermal conductivity of stanene due to their higher group velocities and weaker scattering. Therefore, the orientation-independent phonon group velocities and Gr\"uneisen parameters at low frequencies are crucial to understand the phonon transport in stanene.

\begin{figure}[ht]
\centering
\includegraphics[width=0.45\linewidth]{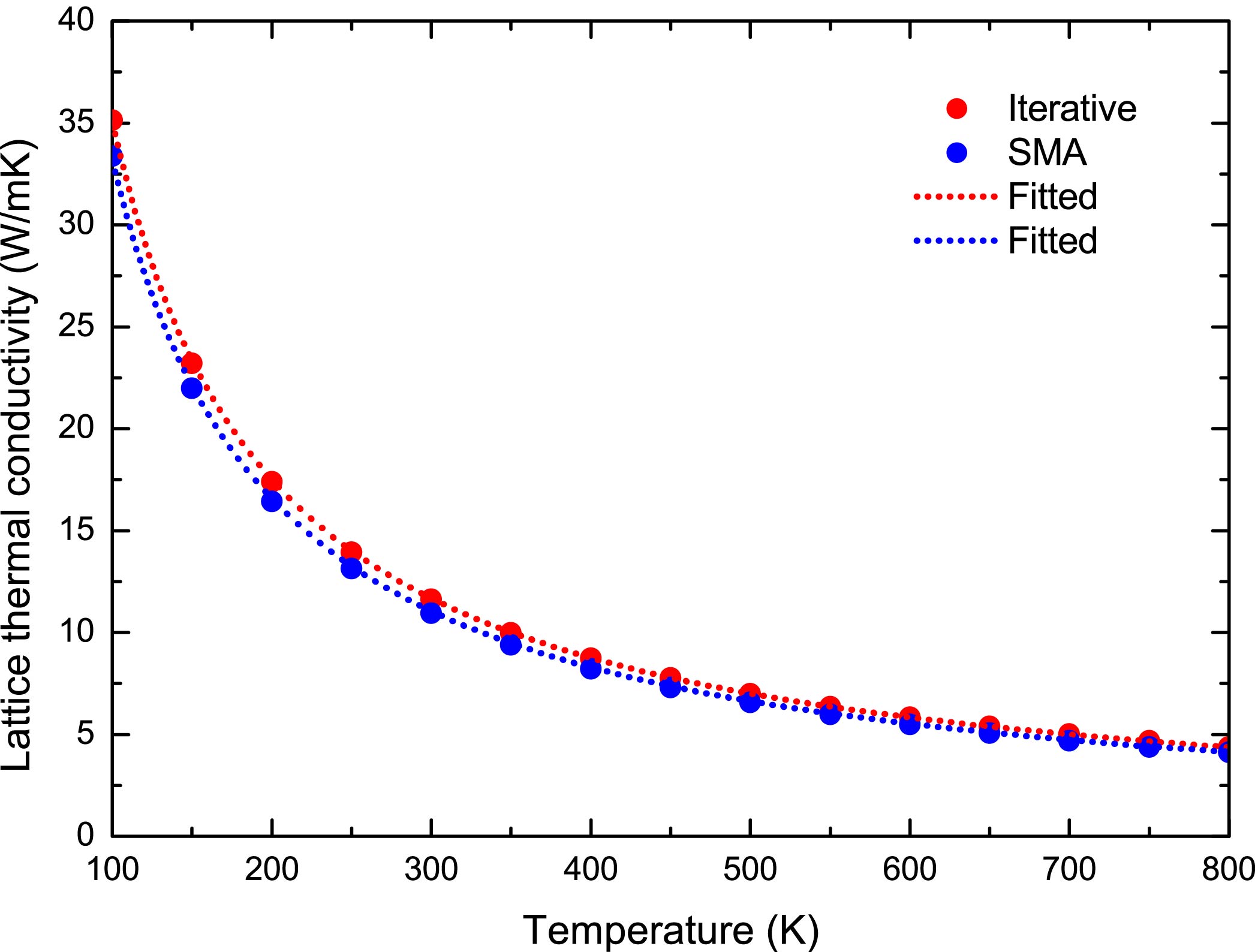}
\caption{Calculated lattice thermal conductivity $\kappa$ of stanene as a function of temperature ranging from 100 K to 800 K using the iterative (red) and SMA (blue) solution of the BTE. The inverse relationship of $\kappa$ and temperature ($\kappa \sim 1/T$) is displayed as dot lines.}
\label{kappa} 
\end{figure}

\subsection{Lattice thermal conductivity}

The intrinsic lattice thermal conductivity $\kappa$ of naturally occurring stanene is calculated using both the iterative and single-mode relaxation time approximation (SMA) solution, as shown in Fig.~\ref{kappa}. In contrast to the anisotropy of thermal conductivity observed in black phosphorene and SnSe \cite{BP-phonon,SnSe,Liu2015}, which possess the hinge-like structure, the isotropic $\kappa$ is found in stanene, which is similar to other 2D hexagonal materials with $C_3$ symmetry such as silicene, blue phosphorene, graphene, and monolayer MoS$_2$ \cite{silicene&graphene,blue-phosphorene,Xu2014,MoS2-SMA,pb2,pb1}.

The thermal conductivity from the iterative solution at 300 K is 11.6 W/mK and 11.9 W/mK for naturally occurring and isotopically pure stanene respectively. Naturally occurring Sn consists of 0.97\% $^{112}$Sn, 0.65\% $^{114}$Sn, 0.36\% $^{115}$Sn, 14.7\% $^{116}$Sn, 7.7\% $^{117}$Sn, 24.3\% $^{118}$Sn, 8.6\% $^{119}$Sn, 32.4\% $^{120}$Sn, 4.6\% $^{122}$Sn, and 5.6\% $^{124}$Sn.

The measured room-temperature $\kappa$ of 500-nm-thick tin film is $46\pm4.2$ W/mK while for the thickness of 100 nm, $\kappa$ is $36\pm2.88$ W/mK \cite{tin-films}, which are lower than the room-temperature $\kappa$ of bulk tin, 64 W/mK. Note that when the thickness of the Sn thin film decreases, phonons become more sensitive to boundary scattering \cite{tin-films}, which further reduces the value of $\kappa$. Thus a smaller calculated $\kappa$ for stanene is a reasonable expectation. The Debye temperature $\Theta_D$ can be calculated from the highest frequency of normal mode vibration (Debye frequency) $\nu_m = 1.51$ THz,

\begin{equation}
\Theta_D=h\nu_m/k_B
\end{equation}

\noindent where $h$ is the Planck constant, and $k_B$ is the Boltzmann constant. The calculated Debye temperature is 72.5 K, which is nearly three times lower than that of $\beta$-Sn of $200\pm3$ K \cite{Bryant1961}. Since the Debye temperature is a measure of the temperature above which all modes begin to be excited and below which modes begin to be frozen out \cite{debye-vibrations,pb2}, this low $\Theta_D$ in stanene also indicates that the $\kappa$ of stanene is much lower than that of bulk tin.

According to the kinetic formula for the thermal conductivity of an assembly of total specific heat $C$, average phonon velocity $v$, and MFP $l$, $i.e.$ $\kappa=1/3 C v l$, when the temperature is well above the Debye temperature $\Theta_D$, the $C$ is constant, while the dominant three-phonon anharmonic processes at high temperatures lead to $l \propto 1/T$ \cite{Ziman,Grimvall}. We fit the calculated $\kappa$ above $\Theta_D$, which reveals an inverse relation to temperature, $i.e.$ $\kappa \sim 1/T$. As shown in Fig.~\ref{kappa}, the fitted $\kappa$ (plotted as dot lines) coincides perfectly with the calculated $\kappa$. Such inverse relationship of $\kappa$ and $T$ is also found in other materials such as graphene, black phosphorene, PbSe, and Mg$_2$(Si,Sn) when $T>\Theta_D$ \cite{selection-rule,BP-phonon,Parker2010,Pulikkotil2012}. Furthermore, we compare the thermal conductivity calculated from the iterative and SMA solution. The difference between the two approaches is very small. It should be noticed that
the SMA assumes that individual phonons are excited independently, which has no memory of the initial phonon distribution, therefore the SMA is inadequate to describe the momentum-conserving character of the Normal processes and it works well only if the Umklapp processes dominate \cite{Fugallo2013}. Our results indicate the dominant role of the Umklapp processes in phonon-phonon interactions over a wide temperature range.

Compared to other 2D hexagonal materials, the $\kappa$ of stanene is smaller than that of silicene (26 W/mK) \cite{silicene&graphene}, monolayer MoS$_2$ ($34.5\pm4$ W/mK) \cite{monoMoS2-k,pb2} and blue phosphorene (78 W/mK) \cite{blue-phosphorene}, while at least two orders of magnitude smaller than that of graphene (3000 W/mK) \cite{silicene&graphene,Xu2014}. The lower $\kappa$ of stanene is due to lower Debye temperature (72.5 K compared to 500 K for monolayer MoS$_2$ and blue phosphorene \cite{MoS2-debye,blue-phosphorene}, and 2,300 K for graphene \cite{Efetov2010}). The Debye temperature reflects the magnitude of sound velocity. Lower Debye temperature results in decreased phonon velocities as mentioned above, and higher phonon scattering rates since more phonon modes are active at a given temperature \cite{debye-vibrations,Lindsay2013,pb2}.

High TE performance requires the system to be a bad conductor for phonons and a good conductor for electrons. As a 2D topological insulator with lower lattice thermal conductivity, stanene can realize much higher thermoelectric efficiency than other 2D materials, which makes it a promising candidate for next-generation thermoelectric devices.

\begin{figure}[ht]
\centering
\includegraphics[width=0.45\linewidth]{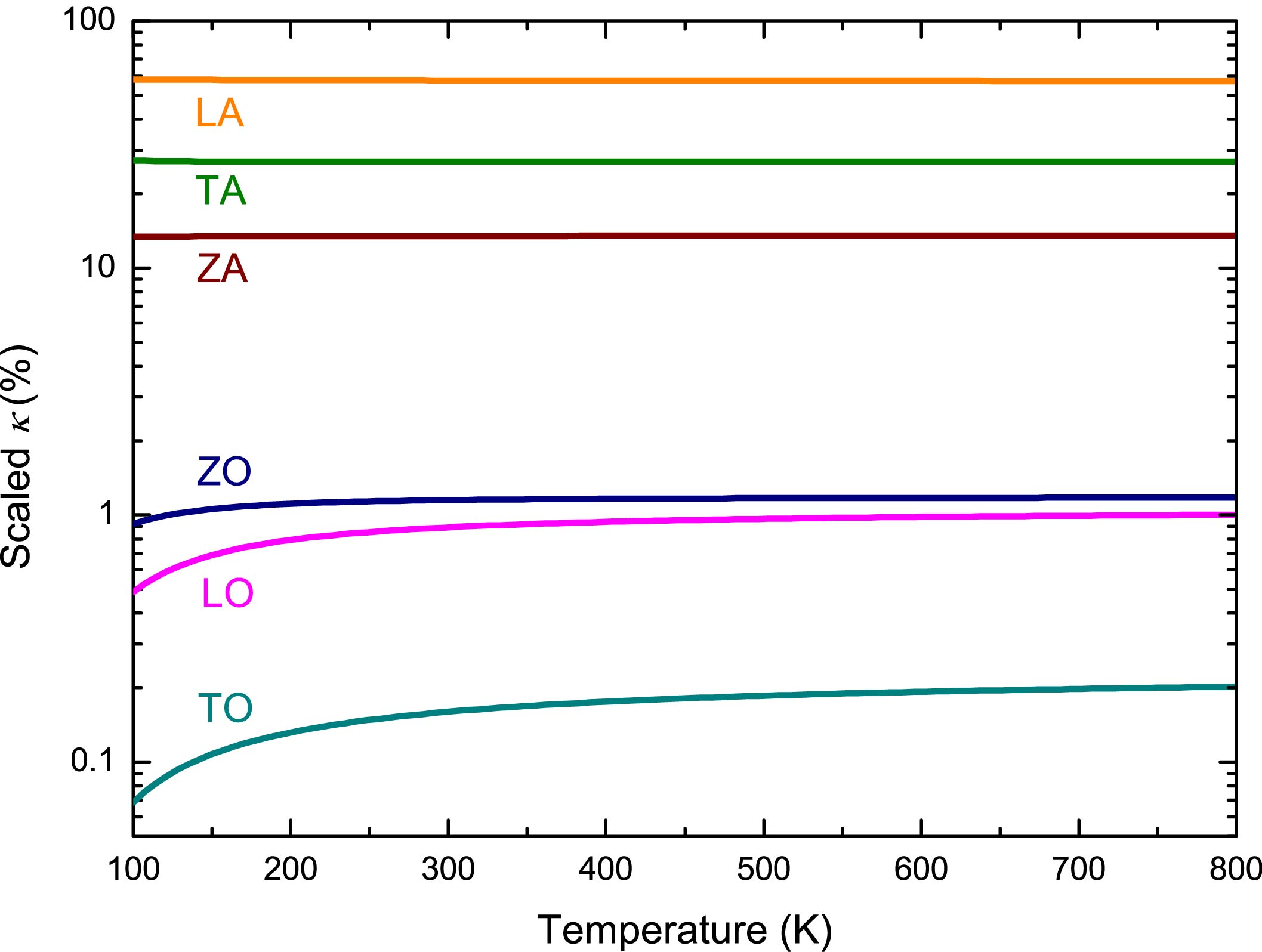}
\caption{Percentage contribution of each phonon mode to $\kappa$ as a function of temperature.}
\label{mode} 
\end{figure}

Furthermore, we examine the contributions of different phonon branches to $\kappa$ of stanene, as shown in Fig.~\ref{mode}. The contribution of LA phonons is more than 57\% at temperatures ranging from 100 K to 800 K, which is larger than 28\% in MoS$_2$ \cite{MoS2-SMA}, 26\% in blue phosphorene \cite{blue-phosphorene} and 8\% in graphene \cite{Lindsay2014}. The contribution of ZA phonons to the $\kappa$ of stanene (13\%) is much smaller than that of graphene (80\%) \cite{silicene&graphene}, blue phosphorene (44\%) \cite{blue-phosphorene}, and monolayer MoS$_2$ (39\%) \cite{MoS2-SMA}, but larger than silicene (7.5\%) \cite{silicene&graphene}. It has been reported that the large contribution of the ZA mode to the $\kappa$ of graphene is due to a symmetry selection rule in one-atom-thick materials, which strongly restricts anharmonic phonon-phonon scattering of the ZA mode \cite{selection-rule}, while the buckled structure of stanene, silicene, and blue phosphorene breaks out the out-of-plane symmetry, in which the selection rule does not apply. As a result, the predicted contribution of ZA phonons to $\kappa$ in stanene is much smaller than that in graphene at 300 K. To provide more physical insight, we investigate the phonon scattering mechanism in detail.

\begin{figure}[ht]
\centering
\includegraphics[width=0.35\linewidth]{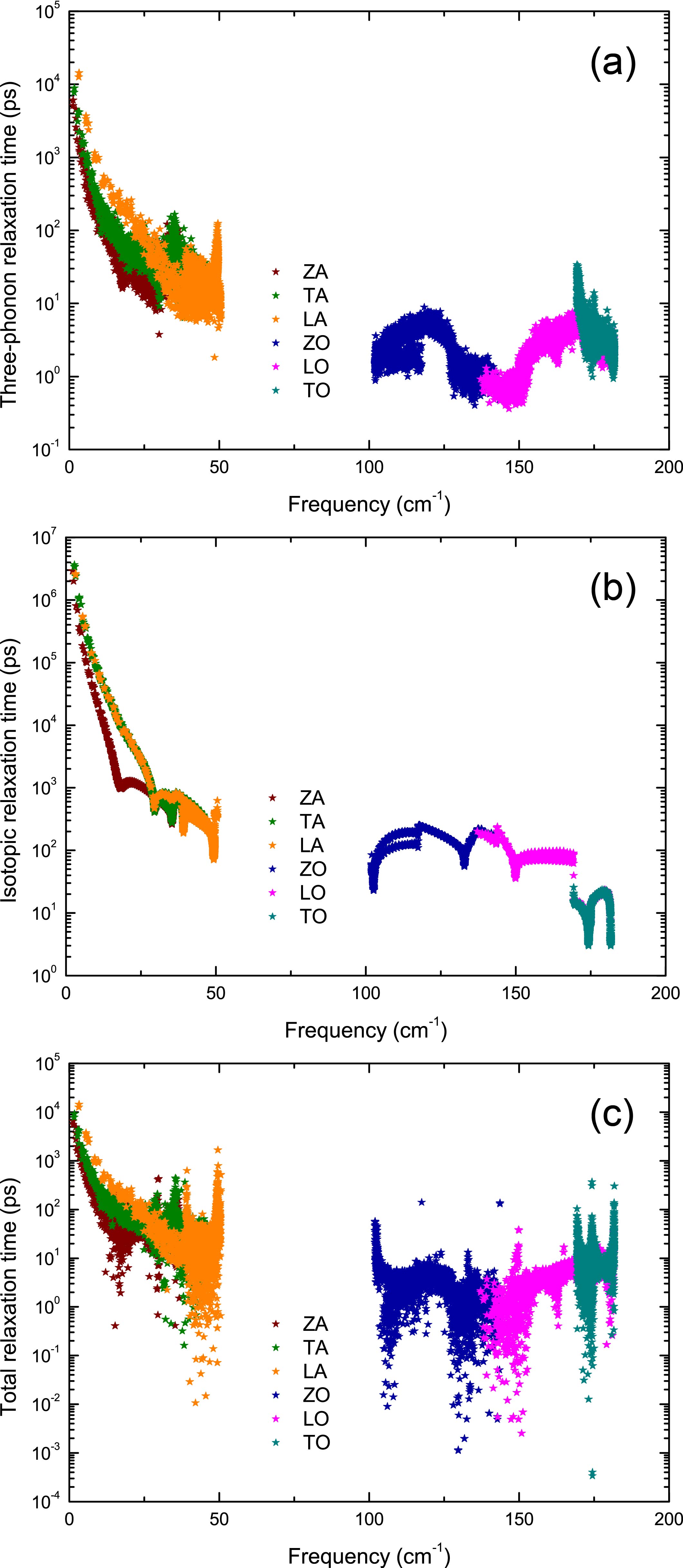}
\caption{The (a)three-phonon, (b)isotopic and (c)total relaxation time of each phonon branch as a function of frequency.}
\label{time} 
\end{figure}

\subsection{Relaxation time and phase space for three-phonon processes}

Considering the significant difference in the contribution of acoustic phonon modes to the $\kappa$ in stanene and other 2D hexagonal materials, it is worthwhile to investigate the relaxation time of each phonon mode as a function of frequency, as shown in Fig.~\ref{time}. In Fig.~\ref{time}(a), the three-phonon relaxation time of LA mode is the longest, and the relaxation time of the optical phonons is much shorter than the acoustic phonons. We also calculate the isotopic relaxation time, as shown in Fig.~\ref{time}(b). Isotopic scattering processes are wavelength-dependent. The long-wavelength phonons can transport all the heat with very little isotopic scattering \cite{Ziman1967}, thus we observe the relative long isotopic relaxation time of acoustic phonons at low frequencies. The inverse of total relaxation time is a sum of contributions from anharmonic three-phonon scattering and isotopic scattering \cite{Ziman,MoS2-SMA},

\begin{equation}
1/\tau_j(\textbf{q})=1/\tau_j(\textbf{q})^{anh}+1/\tau_j(\textbf{q})^{iso}.
\end{equation}

\noindent Thus the three-phonon processes with much shorter relaxation time will dominate the total relaxation time. The total relaxation time of stanene is shown in Fig.~\ref{time}(c). It is in the same order of magnitude as that of graphene, MoS$_2$, and silicene \cite{Lindsay2014,Gu2014,Li2013}. As shown in Fig.~\ref{time}(c), the total relaxation time of LA mode has large values near the acoustic-optical frequency gap, while it is still comparable to that of other acoustic phonon modes at low frequencies. To investigate the phonon scattering mechanism more deeply, we calculate the allowed phase space for three-phonon processes $P_3$.

In the temperature range where three-phonon processes are dominant, the total phase space for three-phonon processes $P_3$ is defined by \cite{phase-space,ShengBTE}

\begin{equation}\label{eqp3}
P_3=\frac{2}{3\Omega}(P_3^{(+)}+\frac{1}{2}P_3^{(-)})
\end{equation}

\noindent where $\Omega$ is a normalization factor, and

\begin{equation}
P_3^{(\pm)}=\sum_j\int{d\textbf{q}D_j^{(\pm)}(\textbf{q})}
\end{equation}

\noindent and 

\begin{equation}\label{eqp5}
D_j^{(\pm)}(\textbf{q})=\sum_{j',j''}\int{d\textbf{q}^\prime \delta(\omega_j(\textbf{q})\pm\omega_{j'}(\textbf{q}^\prime)-\omega_{j''}(\textbf{q}\pm\textbf{q}^\prime-\textbf{G}))}
\end{equation}

\noindent where $D_j^{(+)}(\textbf{q})$ corresponds to absorption processes, $i.e.$ $\omega_j(\textbf{q})+\omega_{j'}(\textbf{q}^\prime)=\omega_{j''}(\textbf{q}+\textbf{q}^\prime-\textbf{G})$, whereas $D_j^{(-)}(\textbf{q})$ corresponds to emission processes, $i.e.$ $\omega_j(\textbf{q})=\omega_{j'}(\textbf{q}^\prime)+\omega_{j''}(\textbf{q}-\textbf{q}^\prime-\textbf{G})$. According to Eq. (\ref{eqp3}-\ref{eqp5}), $P_3$ contains a large amount of scattering events that satisfy the conservation conditions and can be used to assess quantitatively the number of scattering channels available for each phonon mode. Less restricted phase space for three-phonon processes implies a larger number of available scattering channels. Consequently, there is an inverse relationship between $P_3$ and the intrinsic lattice thermal conductivity of a material \cite{Ziman,phase-space}.

\begin{figure}[ht]
\centering
\includegraphics[width=0.35\linewidth]{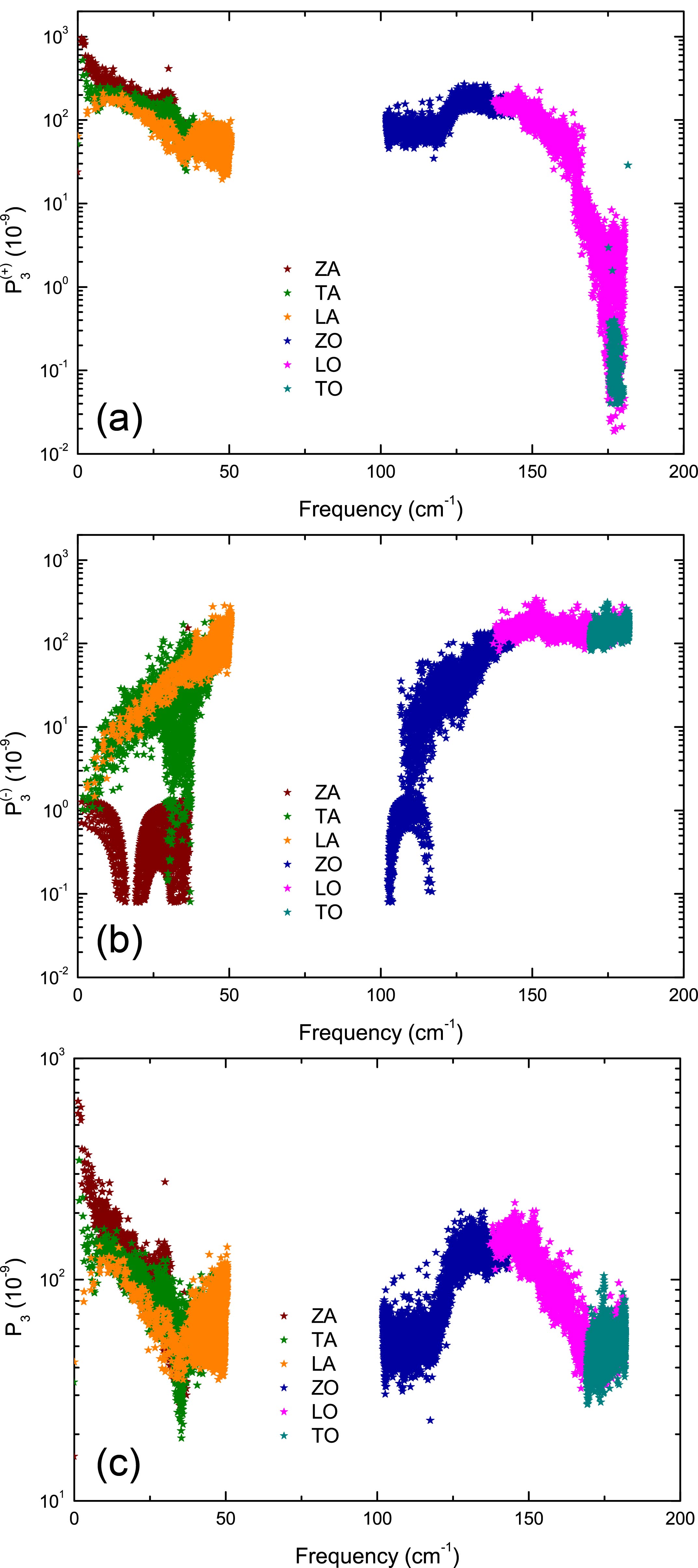}
\caption{Frequency-dependence of three-phonon scattering phase space at 300 K for (a)absorption processes, (b)emission processes and (c)all three-phonon processes.}
\label{p3} 
\end{figure}

The $P_3$ for absorption and emission processes are shown in Fig.~\ref{p3}(a) and (b), respectively. The phase space for absorption processes $P_3^{(+)}$ of LA phonons is much smaller than $P_3^{(+)}$ of ZA and TA phonons, while the phase space for emission processes $P_3^{(-)}$ of LA phonons is comparable to $P_3^{(-)}$ of TA phonons at low frequencies but much larger than that of ZA phonons. According to Eq. (\ref{eqp3}), the calculated total three-phonon phase space $P_3$ for LA phonons is smaller than $P_3$ of ZA/TA phonons in a large part of the low-frequency range, shown in Fig.~\ref{p3}(c). The reason for the relatively small $P_3$ of LA phonons is due to the fact that the energy of LA phonons are relatively high and subsequently the scattering channels, LA+LA/TA/ZA$\leftrightarrow$TA/ZA, are restricted. As a result, the relatively small $P_3$ alongside with relatively large group velocities of LA phonons as mentioned above, determines that LA phonons contribute dominantly to $\kappa$ of stanene.

As for ZA phonons, while the relatively low energy of ZA phonons severely restricts the corresponding phase space for emission processes $P_3^{(-)}$, it also provides a larger number of available scattering channels, ZA+TA/ZA$\leftrightarrow$LA/TA, which leads to larger $P_3^{(+)}$ of ZA phonons, and hence a reduced contribution to $\kappa$ at low frequencies.

It is well known that the dominant contribution of ZA phonons to the $\kappa$ of graphene is due to a symmetry selection rule in such one-atom-thick materials, which strongly restricts anharmonic phonon-phonon scattering of the ZA mode \cite{selection-rule}. The selection rule in graphene that arises from the reflection symmetry perpendicular to the graphene plane restricts the participation of odd number of ZA phonons in three-phonon processes, $e.g.$ ZA+ZA$\leftrightarrow$ZA, ZA+LA/TA$\leftrightarrow$LA/TA. As a result, 60\% of the phase space of ZA phonons is forbidden by the selection rule, which leads to the dominant contribution of the ZA mode to $\kappa$ in graphene. However, for the case of buckled stanene considered here, the reflection symmetry is broken, which means the selection rule does not apply. Consequently, the relatively large $P_3$ of ZA phonons leads to less contribution from ZA phonons to $\kappa$ of stanene, $i.e.$ 13\% in stanene, compared to 80\% in graphene at 300 K.

\subsection{Size dependence of $\kappa$}

\begin{figure}[ht]
\centering
\includegraphics[width=0.9\linewidth]{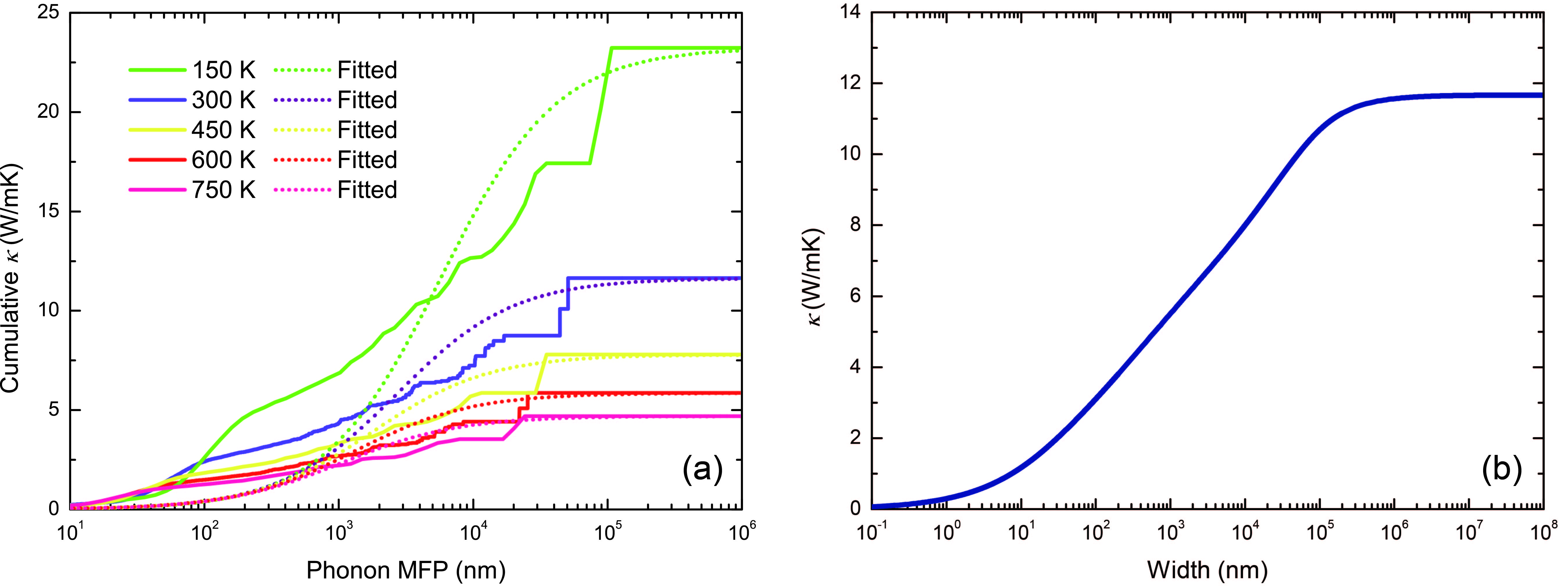}
\caption{(a) Cumulative lattice thermal conductivity of stanene as a function of the phonon MFP at 150 K, 300 K, 450 K, 600 K and 750 K. The curves fitted by Eq. (\ref{eq-cumulative}) are plotted with dot lines. (b) Thermal conductivities of stanene nanowires along [100] direction as a function of width.}
\label{cumulative} 
\end{figure}

Since the thermoelectric figure of merit $zT$ can be improved by optimizing the geometry size to maximize the contribution of the gapless edge states, it is also essential to investigate the phonon transport in stanene nanostructures and achieve the optimized transport properties by reducing thermal conductivity through nanostructuring in the practical design of TE nanodevices. We investigate the size dependence of $\kappa$ by calculating the cumulative thermal conductivity with respect to the maximum MFP allowed. The cumulative thermal conductivity of stanene at different temperatures are plotted in Fig.~\ref{cumulative}(a). The cumulated $\kappa$ keeps increasing as MFP increases, until reaching the thermodynamic limit above a length $L_{diff}$ which represents the longest mean free path of the heat carriers \cite{Fugallo2014,pb1}. The $L_{diff}$ at 150 K, 300 K, 450 K, 600 K and 750 K are 107.2 $\mu$m, 50.9 $\mu$m, 35.1 $\mu$m, 25.5 $\mu$m and 24.2 $\mu$m, respectively. It is found that phonons with MFPs below 73.9 $\mu$m at 150 K, 16.9 $\mu$m at 300 K, 9.5 $\mu$m at 450 K, 8.4 $\mu$m at 600 K and 6.6 $\mu$m at 750 K, respectively, contribute around 75\% of the total $\kappa$, which indicates that the design of nanostructures can be utilized to reduce $\kappa$ and enhance $zT$. In order to properly reduce $\kappa$ for thermoelectric applications, we can fit the cumulative $\kappa$ to a uniparametric function \cite{ShengBTE}

\begin{equation}\label{eq-cumulative}
\kappa(l\leq l_{max})=\frac{\kappa_{max}}{1+\frac{l_0}{l_{max}}},
\end{equation}

\noindent where $\kappa_{max}$ is the ultimate cumulative thermal conductivity, $l_{max}$ is the maximal MFP concerned, and $l_0$ is the parameter to be evaluated. The fitted curves at different temperatures are plotted in Fig.~\ref{cumulative}(a), reproducing both the slope and the position of the calculated data well when $l_{max}>l_0$. It yields the parameter $l_0$ of 5.72 $\mu$m, 2.71 $\mu$m, 1.77 $\mu$m, 1.31 $\mu$m and 0.95 $\mu$m for 150 K, 300 K, 450 K, 600 K and 750 K, respectively, which can be interpreted as representative of the MFP of relevant heat-carrying phonons in stanene. 

Furthermore, when the size gets smaller than $l_0$, the nanostructuring-induced phonon scattering becomes dominant over the three-phonon scattering, and the small-grain-limit reduced $\kappa$ becomes proportional to a constant value $l_{SG}$ \cite{ShengBTE}. We calculate the ratio of the thermal conductivity to the thermal conductivity per unit of MFP in the small-grain limit; for stanene the $l_{SG}$ is found to be 195.0 nm, 92.1 nm, 60.9 nm, 45.6 nm and 36.5 nm at 150 K, 300 K, 450 K, 600 K and 750 K, respectively. This quantity is crucial to the thermal design for modulating the thermal conductivity in the small-grain limit, for example nanowires.

It is worthy to mention that for a nanowire system, phonons with long MFPs will be strongly scattered by the boundary, thus the contribution from which to $\kappa$ will be limited. As shown in Fig.~\ref{cumulative}(b), $\kappa$ decreases with decreasing width of stanene nanowires, and drops to half the maximum $\kappa$ in the thermodynamic limit at widths about 1.3 $\mu$m.

In summary, we predict the lattice thermal conductivity of stanene using first principle calculations and an iterative solution of the BTE. A much lower thermal conductivity is observed in stanene than other 2D materials. Our results indicate that stanene, as a 2D topological insulator with low thermal conductivity, can realize much higher thermoelectric efficiency and is a promising candidate for next-generation thermoelectric devices. The predicted contribution of LA phonons to thermal conductivity in stanene is larger than 57\% due to high phonon group velocities and restricted phase space for three-phonon processes. The representative MFP of stanene is obtained for the purpose of future design in TE-application nanostructures. The MFP in the small-grain limit is also obtained for designing TE devices when the size decreases and the nanostructuring-induced phonon scattering dominates.

\section*{Methods}

All the calculations are performed using the Vienna \textit{ab-initio} simulation package (VASP) based on density functional theory (DFT) \cite{vasp}. We choose the generalized gradient approximation (GGA) in the Perdew-Burke-Ernzerhof (PBE) parametrization for the exchange-correlation functional. We use the projector-augmented-wave potential with 4\textit{d} electrons of tin described as valence, and a plane-wave basis set is employed with kinetic energy cutoff of 500 eV. A 15$\times$15$\times$1 \textbf{k}-mesh is used during structural relaxation for the unit cell until the energy differences are converged within 10$^{-6}$ eV, with a Hellman-Feynman force convergence threshold of 10$^{-4}$ eV/\AA. To eliminate interactions between the adjacent supercells a minimum of 15 \AA\ vacuum spacing is kept.

In the calculation of phonon dispersion, the harmonic interatomic force constants (IFCs) are obtained using density functional perturbation theory (DFPT), which calculates the dynamical matrix through the linear response of electron density \cite{DFPT}. The 5$\times$5$\times$1 supercell with 5$\times$5$\times$1 \textbf{k}-mesh is used to ensure the convergence. The phonon dispersion is obtained using the Phonopy code with the harmonic IFCs as input \cite{phonopy}.

To calculate thermal conductivity of stanene, anharmonic third order IFCs are needed, besides the harmonic IFCs mentioned above. The same 5$\times$5$\times$1 supercell and 5$\times$5$\times$1 \textbf{k}-mesh are used to obtain the anharmonic IFCs, and an interaction range of 8.5 \AA\ is considered herein, which includes fifth-nearest-neighbor atoms. We calculate the thermal conductivity of naturally occuring and isotopically pure stanene by solving the phonon Boltzmann transport equation (BTE) using the ShengBTE code \cite{ShengBTE,ShengBTE1,ShengBTE2}, which is completely parameter-free and based only on the information of the chemical structure.

\section*{Acknowledgements}
This work is supported by the National Natural Science Foundation of China under Grants No. 11374063 and 11404348, and 973 program (No.2013CAB01505).

\end{document}